\documentclass[10pt,a4paper]{iopart}
\pdfoutput=1 

\newcommand{\mean}[1]{\langle #1 \rangle}
\newcommand{\bra}[1]{\left.\langle #1 \right|}

\newcommand{\ket}[1]{\left| #1 \rangle\right.}
\newcommand{\floor}[1]{\lfloor #1 \rfloor}
\newcommand{\dotop}[0]{{\bm\cdot\,}}
\newcommand{\FS}[0]{\rm{FS}}

\newcommand{\vect}[1]{{\mathbf{#1}}}

\newcommand{\vechat}[1]{\hat{\mathbf{ #1}}}

\usepackage{iopams}
\usepackage{mathrsfs}
\usepackage{latexsym}
\usepackage{citesort}
\usepackage{bm}
\usepackage{color}

\usepackage{graphicx}
\usepackage{amsopn}

\begin{document}

\title[Integer effects in entanglement and spin fluctuations]{Integer effects in the entanglement and spin fluctuations of a quantum Hall system with Rashba interactions}
\author{Rona F~Barbarona,${}^{1}$ Francis N~C~Paraan${}^{2}$}
\address{${}^{1}$ Institute of Mathematical Sciences and Physics, University of the Philippines Los Ba\~nos, Laguna 4031, Philippines}
\address{${}^{2}$ National Institute of Physics, University of the Philippines Diliman, Quezon City 1101, Philippines}
\ead{\mailto{rfbarbarona@up.edu.ph}, \mailto{fparaan@nip.upd.edu.ph}}
\date{\today}

\begin{abstract}
We report distinct nonanalytic signatures in the spin-orbit entanglement of a 2D electron gas with Rashba interactions at integer values of the filling factor and at certain level crossings. The accompanying sharp changes in the bulk spin-orbit entanglement entropy can be probed by measuring the fluctuation in the transverse spin polarization of the electron gas. 
\end{abstract}




\hbadness=10000

\section{Introduction}\label{intro}

The Rashba interaction \cite{rashba1960a} models spin-orbit coupling (SOC) that explicitly breaks spatial inversion and time-reversal symmetry. Studies on this type of spin-orbit interaction are important because it can be engineered to manipulate spin-polarized currents in specially designed devices. In particular, controllable spin current polarization is achievable in two-dimensional heterostructure interfaces that are relevant for the fabrication of spintronic devices \cite{datta1990a,slomski2013a} and the realization of topologically-protected states \cite{zhang2010a,qi2011a} that can exhibit the spin Hall effect \cite{sinova2004a,kato2004a,sih2005a,bernevig2006a,valenzuela2006a} and $p$-wave superconductivity \cite{potter2011a}. Additionally, engineering an effective Rashba SOC in ultracold atomic gases in optical traps \cite{galitski2013a} opens up the possibility of studying spin-orbit effects in highly controllable environments. Further applications of the Rashba SOC in the field of spintronics is discussed in a recent review \cite{manchon2015a}.

Additionally, the Rashba SOC leads to entanglement between the spin and orbital degrees of freedom in these systems. This entanglement leads to quantum fluctuations in the transverse spin polarization that persist even in the ground state. In this paper, the first main objective (out of three) is to quantify this spin-orbit entanglement through the von Neumann entropy (\sref{sect:soee}) 
\begin{equation}
	S = -\tr \rho_{\rm{s}} \ln \rho_{\rm{s}},
\end{equation}
that is obtained from the reduced density operator on the spin subspace
\begin{equation}
	\rho_{\rm{s}} = \tr_{\rm{o}} \ket{\FS}_{{\rm{so}}} \bra{\FS}_{{\rm{so}}}.
\end{equation}
Here $\ket{\FS}_{{\rm{so}}}$ refers to the many-electron ground state and the partial trace is done over state vectors in the orbital subspace. We refer to this entanglement entropy as the spin-orbit entanglement entropy (SOEE). Spin-orbit entanglement due to the Rashba SOC has been previously studied for a geometrically confined quasi-one-dimensional electron system, but it has only been calculated for a single electron \cite{safaiee2009a,safaiee2011a}. {Additionally, the generation of spin-orbit entanglement in single-electron wavepackets traversing mesoscopic 2D slabs with Rashba interactions has been analyzed numerically to investigate spin accumulation along the slab edges \cite{nikolic2005a}. This type of entanglement between spin and orbital degrees of freedom, which has also been analyzed in quantum electrodynamics devices \cite{deng2015a}, is distinct from the block entanglement that is present in quantum Hall states partitioned in real space \cite{rodriguez2009a,petrescu2014a}. The latter entanglement arises from electron-electron interactions and leads to spatial correlations between block partitions and charge fluctuations within them.} 

In clean quantum Hall geometries the Rashba SOC modifies the electronic band structure of the many-electron system, but maintains the flatness of the bands of delocalized states (\sref{model}). The density of delocalized states thus remains sharp about the band centers, which causes several thermodynamic and transport quantities of the 2D electron gas to exhibit pronounced changes as the Fermi level jumps discontinuously between delocalized band gaps \cite{peierls1933a,bychkov1983a,bychkov1984a,shen2004a,wang2009a,gammag2012a}. Indeed, experimental observations of the Shubnikov--de Haas effect and de Haas--van Alphen effect have been used to obtain the value of the Rashba parameter \cite{bychkov1983a,das1989a,nitta1997a,mineev2005a,wang2009a,gammag2012b,gammag2013a,bell2013a,rupprecht2013a,wilde2009a,wilde2013a,wilde2014a}. The second objective of this work is to show that similar integer effects exist in the total spin-orbit entanglement entropy originating from the Rashba interaction (\sref{sect:soee}). These nonanalytic features in bulk entanglement measures take the form of kinks (discontinuous first derivatives with respect to the external field) at certain values of the filling factor. 

The presence of spin-orbit entanglement further means that spin measurements on the reduced spin state are statistically distributed according to the reduced density operator $\rho_{\rm{s}}$. Finally, our third objective (\sref{sect:polarization}) is to calculate the average spin polarization along the transverse axis and the accompanying polarization fluctuations. We report clear signatures in the average transverse polarization of crossings between the highest occupied energy level and the lowest unoccupied level at special values of the filling factor. Furthermore, since subsystem fluctuations in bipartitioned systems are indicators of entanglement \cite{song2012a,calabrese2012a,villaruel2016a} we demonstrate that the integer effects observed in the SOEE are also manifested in the polarization fluctuations.

\section{Hamiltonian}\label{model}
The model is a two-dimensional electron gas that is confined along a constant $z$ plane and immersed in a uniform external magnetic field $\vect{B} = B\vechat{z} = \bm\nabla \times \vect{A}$. {We take an independent electron approach that allows an exact calculation of the bulk SOEE while maintaining a tractable description of spin-orbit interactions. The single-electron Hamiltonian considered here adds to the Landau model of the quantum Hall effect a Rashba spin-orbit coupling term of strength $\alpha$ \cite{rashba1960a}}: 
\begin{equation}\label{rashbaH}
H = \frac{(\hbar\vect{k} + e\vect{A})^2}{2m}\,\sigma_0 + \frac{1}{2} g\mu_{\rm{B}}\vect{B}\, \dotop \bm\sigma+ \frac{\alpha}{\hbar}\Bigl[(\hbar\vect{k}+e\vect{A})\times\bm\sigma \Bigr]\dotop\vechat{z}.
\end{equation}
In this Hamiltonian $m$ is the effective mass, $e$ is the elementary charge, $g$ is the gyromagnetic factor, $\mu_{\rm{B}}$ is the Bohr magneton, and the Pauli operators $\sigma_\mu$ act on the electron spin space ($\sigma_0$ is the identity operator). Periodic boundary conditions are imposed and we choose the symmetric gauge $\vect{A} = \frac{1}{2}(-y\hat{\vect{x} }+ x \hat{\vect{y}})$. In the absence of Rashba SOC ($\alpha=0$) the spin-resolved single-electron states $\ket{\tilde{n},\uparrow\downarrow}$ are labeled by the Landau level index $\tilde{n}$ and the spin index $\uparrow\downarrow$. If $A$ is the area of the sample and $\omega_{\rm{c}} = eB/m$ the cyclotron frequency, the state $\ket{\tilde{n},\uparrow\downarrow}$ has degeneracy $d= A m \omega_{\rm{c}}/(2\pi \hbar)$. 

For non-zero $\alpha$ the eigenvectors and energy eigenvalues of this Hamiltonian have been calculated \cite{das1990a,schliemann2008a}. To diagonalize the Rashba Hamiltonian \eref{rashbaH} we first make the transformation to canonical boson operators $[b,b^\dag] =1$:
\begin{eqnarray}
b^\dag = \frac{1}{\sqrt{2}}\frac{\ell}{\hbar}[(\hbar k_y + e A_y) - \rmi (\hbar k_x+ e A_x) ],\\
b = \frac{1}{\sqrt{2}}\frac{\ell}{\hbar}[(\hbar k_y + e A_y) + \rmi (\hbar k_x+ e A_x) ].
\end{eqnarray}
Here the magnetic length is $\ell = \sqrt{\hbar/(eB)}$ and the Hamiltonian becomes
\begin{equation}
H = \hbar\omega_{\rm{c}} \bigl(b^\dag b + \case{1}{2}\bigr)\sigma_0 + \hbar\omega_{\rm{c}} \xi\sigma_z + \frac{\alpha\sqrt{2}}{\ell}\bigl(b\sigma_- + b^\dag\sigma_+\bigr),
\end{equation}
with dimensionless constant $\xi = g\mu_{\rm B} m/(2\hbar e)$ and spin lowering/raising operators $\sigma_\pm = \frac{1}{2}(\sigma_x \pm \rmi\sigma_y)$. In this form, the Hamiltonian is block diagonal in the Landau level basis $\ket{\tilde{n},\uparrow\downarrow}$ and is equivalent to the Jaynes-Cummings model of atom-field interactions after a spin-flip transformation \cite{jaynes1963a,schliemann2008a,bouziani2012a}. Thus, the result for the atom-field entanglement entropy reported in the Jaynes--Cummings model \cite{phoenix1991a} is similar to the entanglement entropy for a single electron described by the Rashba Hamiltonian \eref{rashbaH}.

Continuing with the diagonalization, we obtain the following energy eigenvalues:
\begin{eqnarray}
E_{0,+}= \hbar \omega_{\rm{c}}(1+2\xi)/2,\\
E_{n\sigma} = \hbar \omega_{\rm{c}} \bigl[n + \sigma\case{1}{2}\sqrt{\gamma n + (1+2\xi)^2} \bigr], \quad(n=1,2,\cdots;\ \sigma = \pm 1), 
\end{eqnarray}
with the dimensionless quantity $\gamma = 8m\alpha^2/(\hbar^3\omega_{\rm{c}})$ that measures the relative strength of the Rashba interaction and the bare Landau level splitting $\hbar\omega_{\rm{c}}$. These eigenvalues correspond to the energy eigenstates
\begin{eqnarray}
\ket{0,+1}&\equiv \ket{0,\uparrow},\\
\ket{n,\sigma}&\equiv u_{n\sigma}\ket{{n},\uparrow} + v_{n\sigma}\ket{{n}-1,\downarrow}, 
\label{rashbastates}
\end{eqnarray}
which are labeled by the Rashba index $n$ and a polarity index $\sigma$. Thus, single-electron eigenstates are linear combinations of non-interacting states belonging to adjacent Landau levels with opposite spin index, except for the state $\ket{0,+1}$, which is unaffected by the Rashba SOC. The squares of the expansion coefficients are given by
\begin{equation}\label{eq:prob}
\left|u_{n\sigma}\right|^2 = \frac{1}{2}\Biggl[1 + \sigma \, \frac{1+2\xi}{\sqrt{\gamma n + (1+2\xi)^2}}\Biggr] = 1- \left| v_{n\sigma} \right|^2.
\end{equation}

The band structure of the many-electron 2DEG with Rashba interactions consists of $d$-fold degenerate single-electron states within each Rashba energy level $\ket{n,\sigma}$. In the $\ket{\FS}$  ground state with filling factor $\nu$, the $\lfloor{\nu}\rfloor$ levels with lowest energies are filled and the next energy level is partially filled. When the filling changes with applied field $B = N_a h/(\nu e)$, the Fermi level jumps discontinuously at integer filling factors. As we discuss in the following sections, this discontinuous jump can lead to nonanalytic behavior in the entanglement and transverse polarization fluctuations in the ground state of a 2DEG with Rashba interactions.

\section{Spin-orbit entanglement entropy}\label{sect:soee}
The eigenstates $\ket{n,\sigma}$ are Schmidt decomposed \cite{ekert1995a} with respect to a partitioning between the spin degrees of freedom (labeled by $\uparrow\downarrow$) and orbital degrees of freedom (labeled by the Landau index $\tilde{n}$). The reduced density operator on the spin subspace of the pure state $\ket{n,\sigma}\negthinspace\negthinspace{\bra{n,\sigma}}$ is therefore
\begin{equation}
\rho_{\rm{s}}(n,\sigma) = \tr_{\tilde{n}} \ket{n,\sigma}\negthinspace\negthinspace\bra{n,\sigma} =\left|{u_{n\sigma}}\right|^2 \left| \uparrow \rangle\right. \negthinspace\negthinspace\bra{\uparrow} + \left|v_{n\sigma}\right|^2 \ket{\downarrow}\negthinspace\negthinspace\bra{\downarrow},
\end{equation}
and the partial SOEE associated with an occupied $\ket{n,\sigma}$ state is \cite{safaiee2009a,safaiee2011a}
\begin{equation}\label{partialsoee}
S_{n\sigma} = -\left|u_{n\sigma}\right|^2 \ln \left|u_{n\sigma}\right|^2 - \left|v_{n\sigma}\right|^2 \ln \left|v_{n\sigma}\right|^2.
\end{equation}
We are interested in the total SOEE in the many-electron ground state $\ket{\FS}$ in which $\nu d$ single-electron states below the Fermi energy are occupied. Since the von Neumann entanglement entropy is additive over independent systems, the SOEE per electron is
\begin{equation}\label{save}
s = \frac{1}{N_a}\frac{m\omega_{\rm c}}{2\pi\hbar} \biggl[(\nu -\floor{\nu}) S_{M\Sigma} + \sum_{\{m,\sigma\}} S_{m\sigma}\biggr],
\end{equation}
where $N_a$ is the area density of electrons. The set $\{m,\sigma\}$ contains the quantum numbers of the $\floor{\nu}$ fully occupied Rashba levels below the Fermi energy, while $M,\Sigma$ are the quantum numbers of the highest partially occupied Rashba level.

\begin{figure}[tb]
	\centering
		\includegraphics[width=0.48\linewidth]{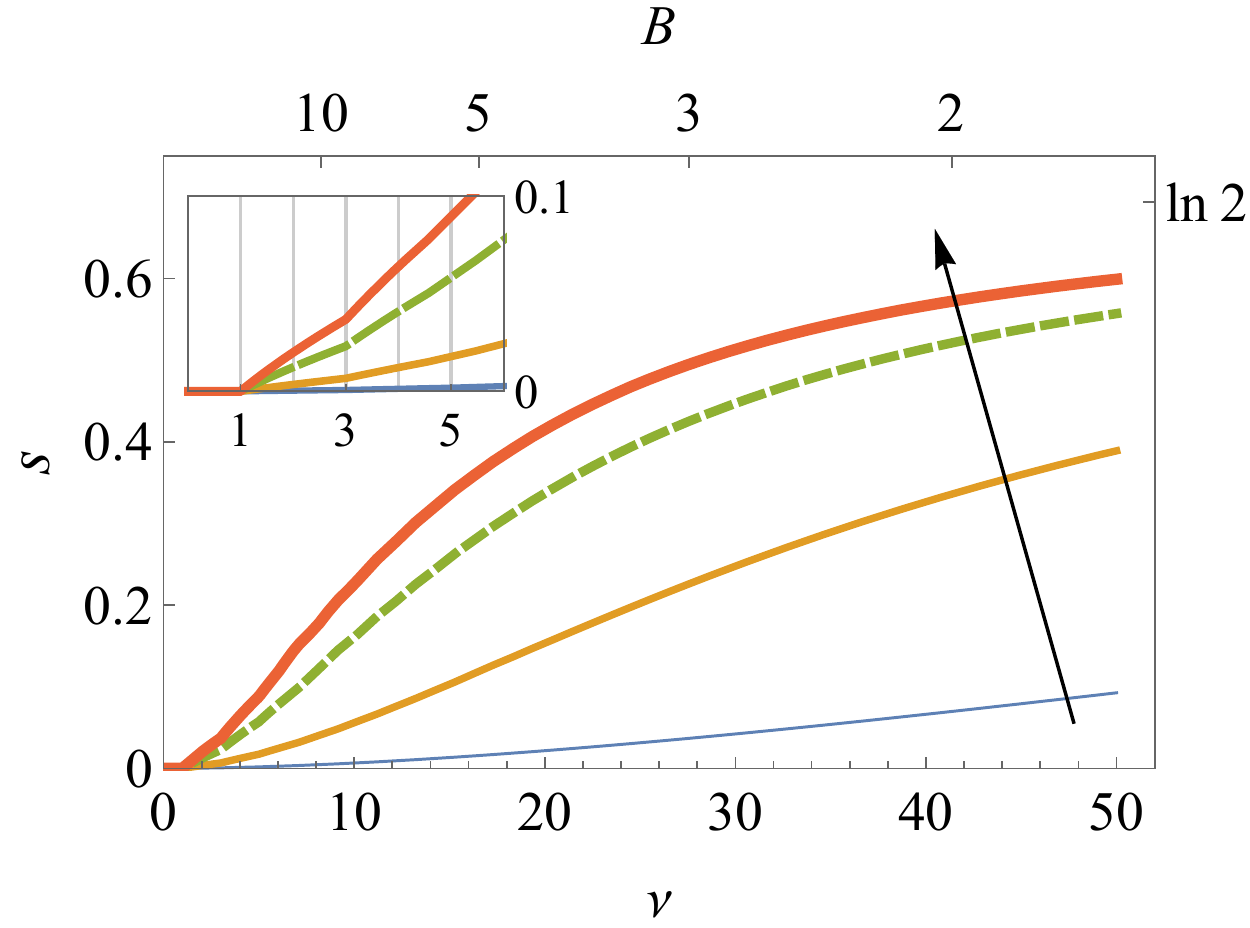}\hfill \includegraphics[width=0.48\linewidth]{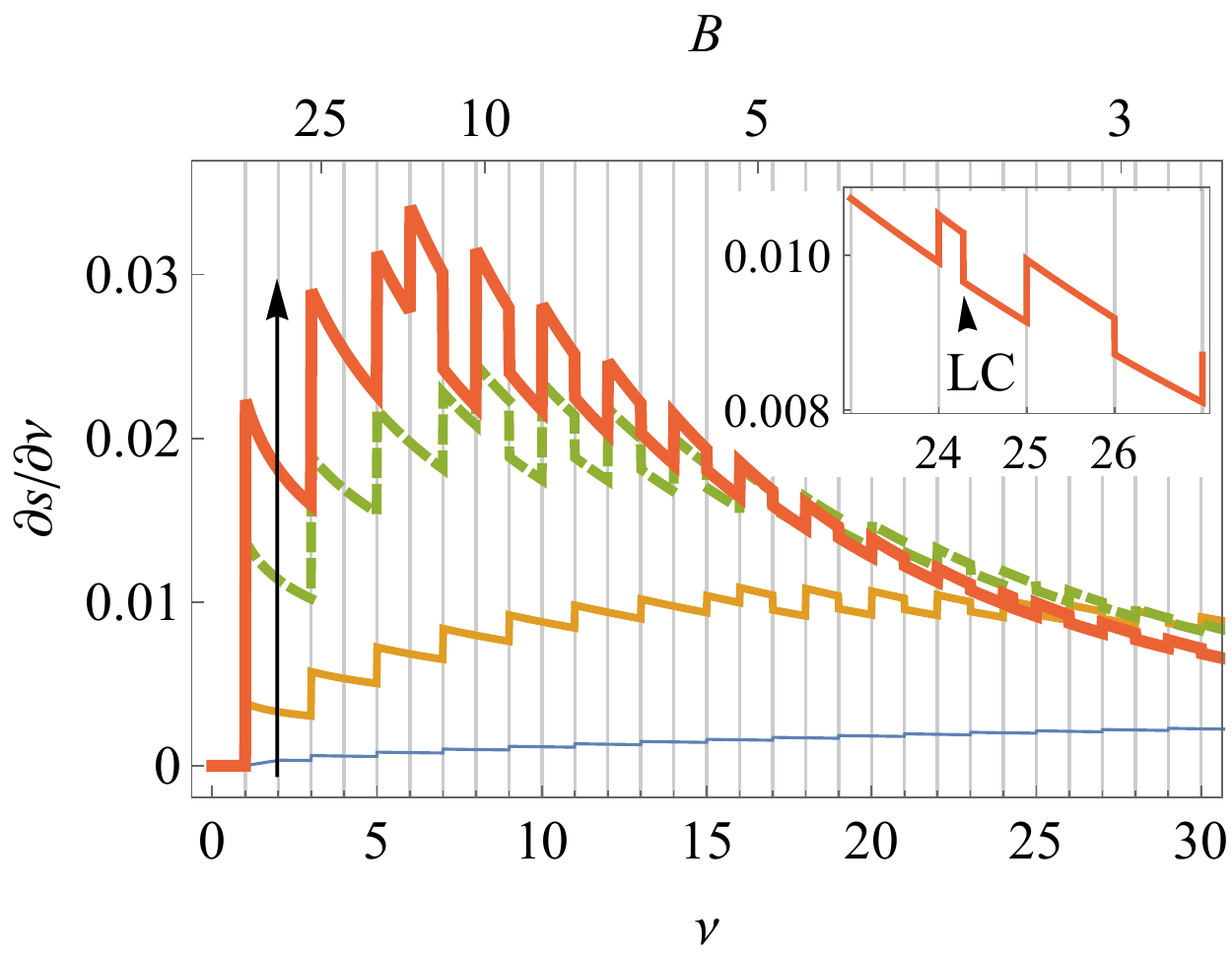}
	\caption{The spin-orbit entanglement entropy per electron $s(\nu)$ is piecewise smooth and saturates to $\ln 2$ at large filling factors (left panel). It has discontinuous derivatives at some integer values of $\nu$ as Rashba levels with different $n$ are filled in succession (right panel). Level crossings about the highest occupied energy level can also lead to nonanalyticities in the SOEE, as depicted in the inset where $\alpha \times 10^{12} = 20e$ and $\nu = 24.28$ (LC). The Rashba parameter $\alpha$ increases in the direction of the black arrows with $\alpha\times 10^{12} = 2.0e, 7.2e, 15e, 20e$. Other parameters have been chosen to be consistent with a 2DEG in InGaAs heterostructures ($m = 0.05 m_e, g = -4, N_a = 2 \times 10^{16}\, \rm{m}^{-2}$) \cite{cangas2009a}.}\label{fig:soee}
\end{figure}

In~\fref{fig:soee} the SOEE per electron $s$ is plotted as a function of the filling factor for several values of the Rashba parameter. As the filling factor is increased more and more energy levels with large quantum number $n$ become filled and the average SOEE $s$ approaches the limiting value $\ln 2$. This maximally entangled situation occurs because $\lim_{n\to\infty} \left|u_{n\sigma}\right|^2 = \lim_{n\to\infty} \left|v_{n\sigma}\right|^2  = 1/2$ and at large $\nu$ the reduced state $\rho_{\rm{s}}$ increasingly becomes an equal mixture of electrons in the $\uparrow$ and $\downarrow$ states. 

Close examination of~\fref{fig:soee} reveals that $s(\nu)$ has several kinks at certain values of the filling factor. To visualize the kinks in the SOEE per electron better, the derivative $\partial s/\partial \nu$ is also graphed, which clearly shows the location of discontinuous derivatives. {The origin of this nonanalytic behavior can be seen from the exact result \eref{save}: As the filling factor $\nu$ is increased, the fractional occupation of the highest Rashba level (with quantum numbers $M$ and $\Sigma$) also increases and the rate of change in $s(\nu)$ depends on the level degeneracy and the partial SOEE $S_{M\Sigma}$. The latter is a discontinuous function of the integer Rashba index $M$ \eref{partialsoee} and thus the SOEE per electron $s(\nu)$ is nonanalytic at filling factors that satisfy the following conditions. First, kinks in $s(\nu)$ occur at some integer values of $\nu$ where the Rashba quantum number $M$ of the highest occupied level changes as the Fermi energy jumps across gapped Rashba levels.} At high filling and large $\alpha$, this situation happens at every integer $\nu$. However, at low filling and weak Rashba SOC, the discontinuities in $\partial s /\partial \nu$ are not observed at even integer filling factors where only the polarity index $\sigma$ of the highest occupied level changes (since $S_{M,\Sigma} = S_{M,-\Sigma}$). The second instance where $s(\nu)$ is nonanalytic occurs at a level crossing between a highest occupied level and a lowest unoccupied level with different Rashba indices. This latter situation explains discontinuous derivatives at noninteger $\nu$, such as the example LC shown in the inset of \fref{fig:soee}.

\begin{figure}[tb]
	\centering
		\includegraphics[width=0.48\linewidth]{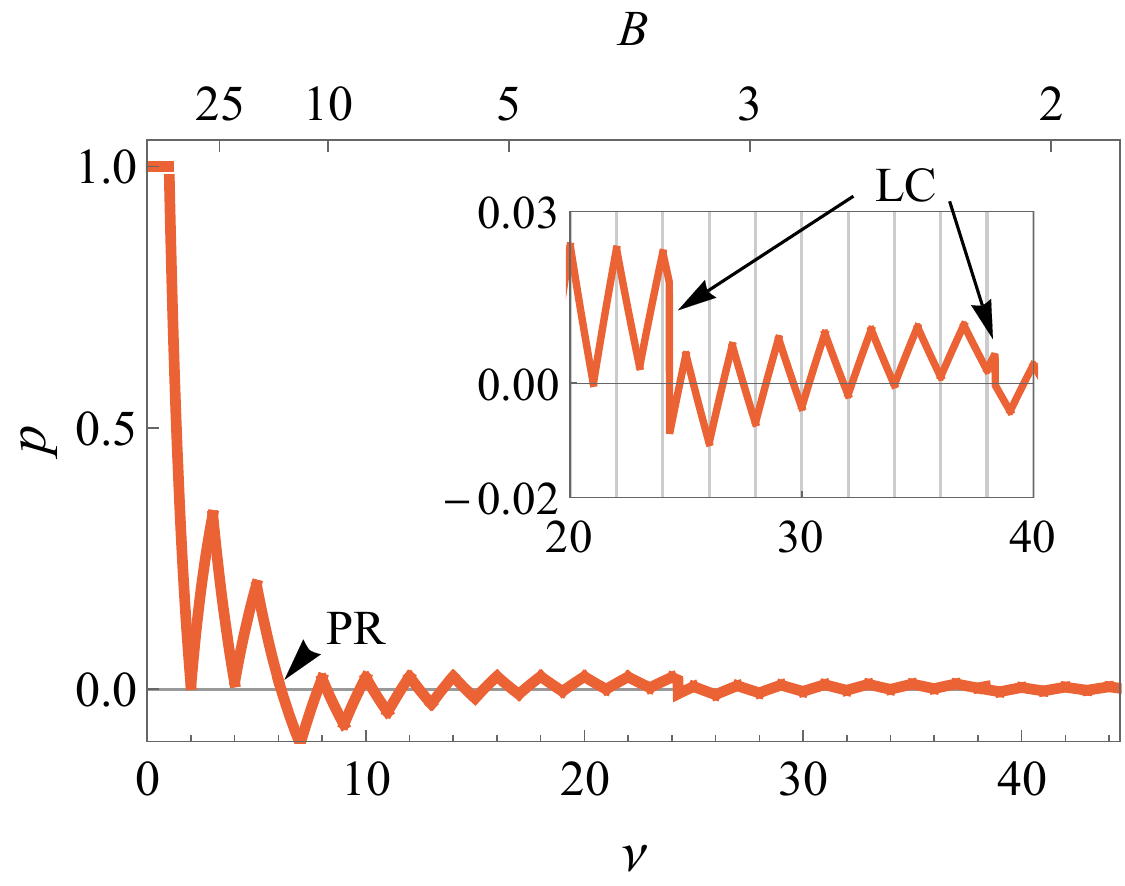}\hfill \includegraphics[width=0.48\linewidth]{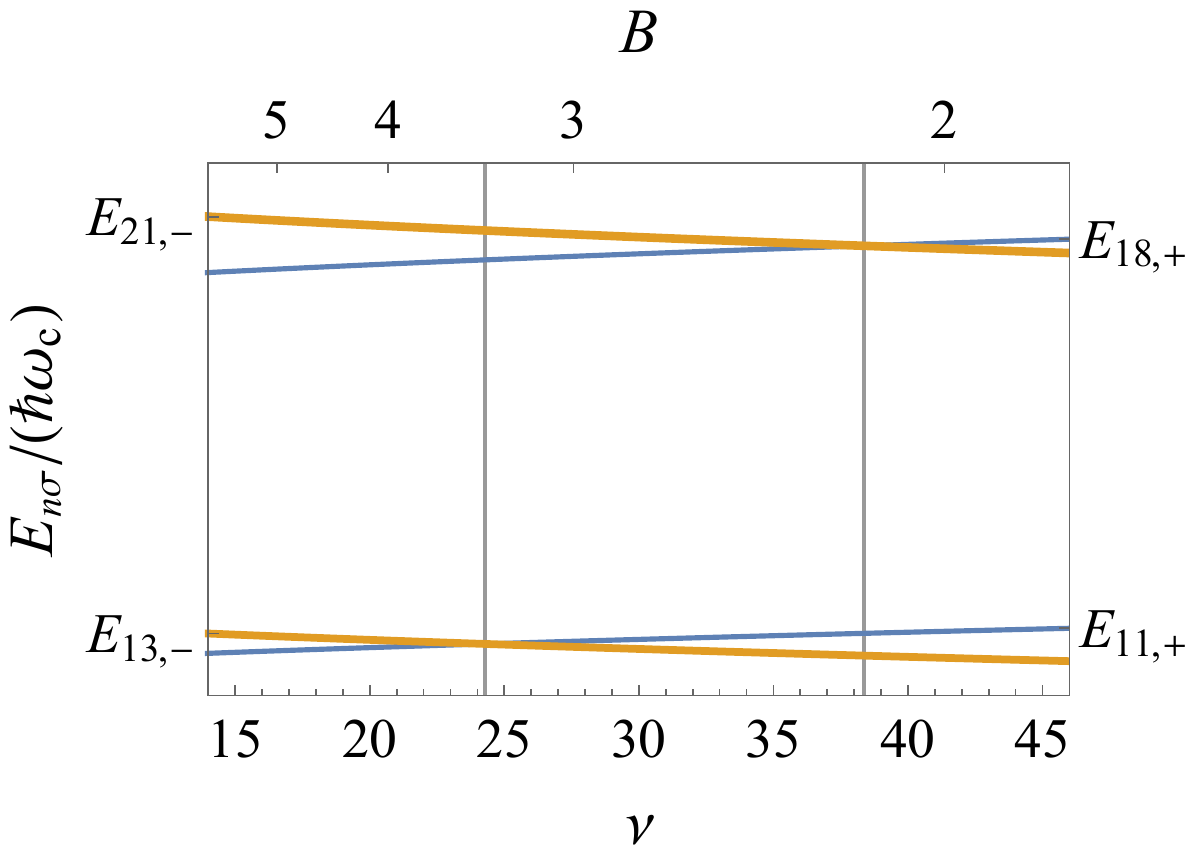}
	\caption{The transverse polarization per electron $p$ oscillates when the Fermi level crosses levels of opposite polarity $\sigma$ at integer filling $\nu$ (left panel). Polarization reversals (PR) can occur. Discontinuities in $p$ arise at level crossings about the Fermi level (LC). Such crossings between the highest occupied and lowest unoccupied levels occur at  $\nu = 24.28$ and $\nu=38.36$ in this example (right panel). In these plots $\alpha\times 10^{12} = 20e$ and other parameters have been chosen to be consistent with a 2DEG in InGaAs heterostructures ($m = 0.05 m_e, g = -4, N_a = 2 \times 10^{16}\, \rm{m}^{-2}$) \cite{cangas2009a}.}\label{fig:polarization}
\end{figure}

\section{Transverse spin polarization and fluctuations}\label{sect:polarization}
Due to the spin-orbit interaction the Pauli operator $\sigma_z$ fails to commute with the Hamiltonian and the polarization about the transverse $z$-axis is no longer a good quantum number. The transverse spin polarization of an electron occupying an $\ket{n,\sigma}$ state is {$\mean{P_{n\sigma}} \equiv \bra{n,\sigma}\sigma_z\ket{n,\sigma}= \left|u_{n\sigma}\right|^2 - \left|v_{n\sigma}\right|^2$} so that the average transverse polarization per electron $p$ of the ground state is
\begin{equation}
p =  \frac{1}{N_a}\frac{m\omega_{\rm c}}{2\pi\hbar} \biggl[(\nu -\floor{\nu}) \mean{P_{M\Sigma}} + \sum_{\{m,\sigma\}} \mean{P_{m\sigma}}\biggr].  \label{eq:pol}
\end{equation}
The transverse polarization $p$ is a quantum expectation value at zero temperature, in contrast to the thermodynamic magnetization obtained from the field derivative of a thermodynamic potential \cite{lifshitz1956a}. Thus, the polarization oscillations depicted in \fref{fig:polarization} are different from the magnetization oscillations that are associated with the de Haas--van Alphen effect. In the present example, the polarization $p$ has kinks at integer $\nu$ and oscillations occur when the polarities $\sigma$ of the highest occupied level change. In some cases the alternating pattern in the polarization is broken, which can result in a reversal of spin polarization (labeled PR in \fref{fig:polarization}). 

\begin{figure}[tb]
\centering
	\includegraphics[width=.55\linewidth]{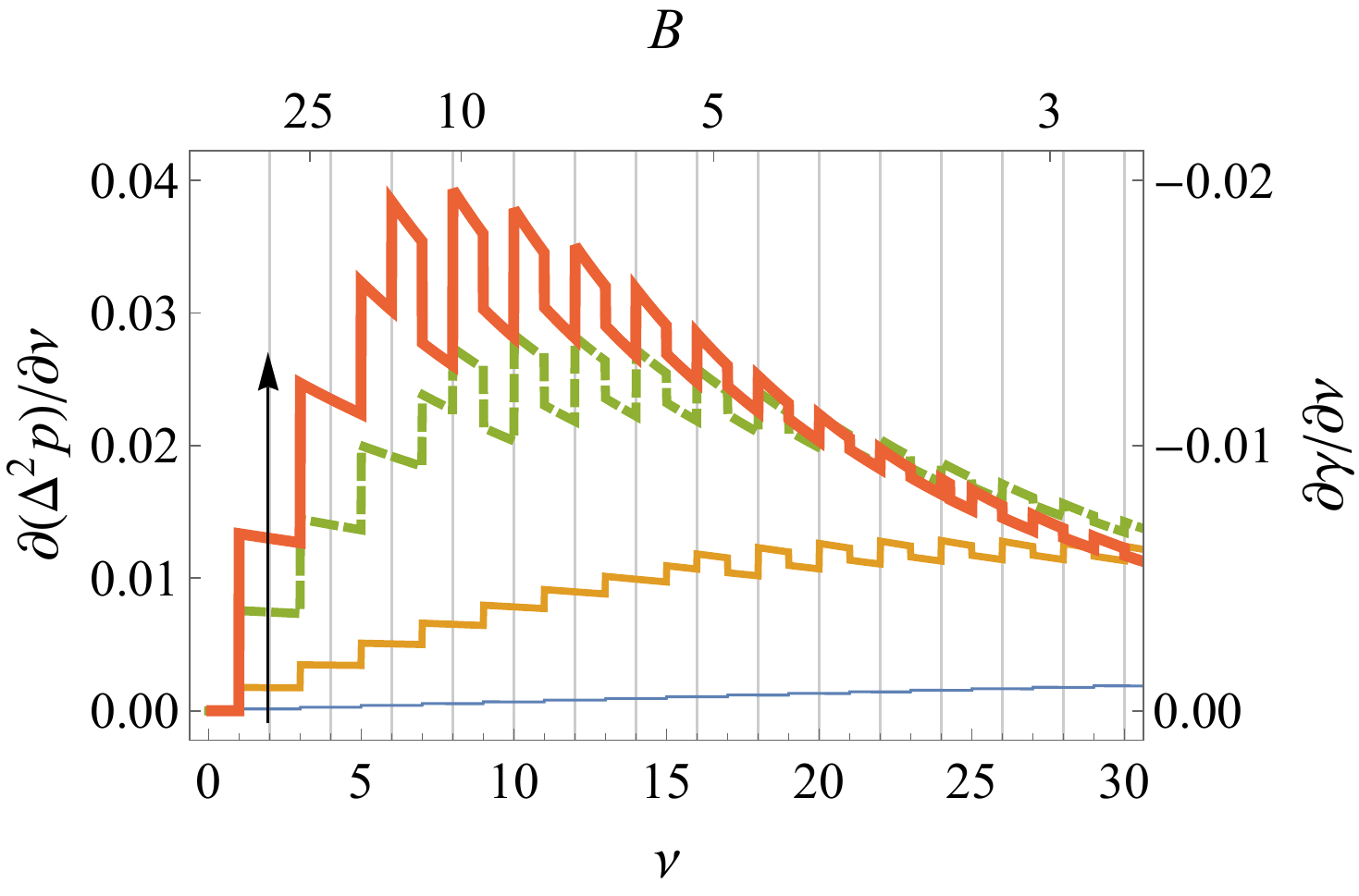}
\caption{The transverse polarization fluctuation per electron $\Delta^2 p(\nu)$ and quantum purity per electron $\gamma$ have discontinuous derivatives at the same values of $\nu$. The Rashba parameter $\alpha$ increases in the direction of the black arrow with $\alpha\times 10^{12} = 2.0e, 7.2e, 15e, 20e$. Other parameters have been chosen to be consistent with a 2DEG in InGaAs heterostructures ($m = 0.05 m_e, g = -4, N_a = 2 \times 10^{16}\, \rm{m}^{-2}$) \cite{cangas2009a}.}\label{fig:polfluc}
\end{figure}

As mentioned earlier, the presence of spin-orbit entanglement leads to uncertainty in determining the spin state of the electrons in the 2DEG. The fluctuation or variance in the transverse spin polarization per electron in the 2DEG is given by
\begin{equation}
\Delta^2 p \equiv  \frac{4}{N_a}\frac{m\omega_{\rm c}}{2\pi\hbar}\biggl[(\nu -\floor{\nu}) \left|u_{M\Sigma}\right|^2 \left|v_{M\Sigma}\right|^2 + \sum_{\{m,\sigma\}} \left|u_{m\sigma}\right|^2 \left|v_{m\sigma}\right|^2\biggr]. \label{eq:sigma2}
\end{equation}
As seen in \fref{fig:polfluc} the derivative $\partial(\Delta^2 p)/\partial \nu$ has the same qualitative features as $\partial s/\partial \nu$. That is, the transverse polarization fluctuation and the entanglement entropy exhibit similar nonanalytic behavior at the filling factors $\nu$ where the Rashba index of the highest occupied energy level changes. 

An exact quantitative relationship between these fluctuations and the spin-orbit entanglement can be made by applying an entanglement-fluctuation relation \cite{villaruel2016a} for each occupied single-electron state. It turns out that the polarization fluctuation for an electron in the state $\ket{m,\sigma}$ is directly related to entanglement measures such as the purity $\gamma_{m\sigma} = \tr\, [\rho_{\rm{s}}(m,\sigma)]^2$ and order-2 R\'enyi entropy $R_{m\sigma} \equiv - \ln \gamma_{m\sigma}$:
\begin{equation}
	\mean{\Delta^2p_{m\sigma}} = 4 \left|u_{m\sigma}\right|^2 \left|v_{m\sigma}\right|^2 = 2(1 - \gamma_{m\sigma}) = 2 (1-\rme^{-R_{m\sigma}}).
\end{equation}
If we define $\gamma$ as the average purity per electron we obtain $\Delta^2p = 2(1-\gamma)$ and find that the nonanalyticities in spin-orbit entanglement are mirrored by those in the transverse polarization fluctuation  (\fref{fig:polfluc}). Thus, such singular features of the spin-orbit entanglement can, in principle, be observed by spin polarization measurements on a Rashba coupled 2DEG. The challenge of isolating these ground state polarization fluctuations from thermal ones may be more easily addressed in the analogous ultracold atom setting \cite{galitski2013a} than in a heterostructure interface.

\section{Concluding remarks}	

Spin-orbit entanglement arising from the Rashba interaction leads to ground state fluctuations in the transverse spin polarization of a 2D electron gas. In this paper, we have described new nonanalytic features in the bulk spin-orbit entanglement entropy of such a gas at integer values of the filling factor and at level crossings. With the use of an entanglement-fluctuation relation, we have argued that such discontinuous features of the spin-orbit entanglement can be observed by determining the variance in measurements of the transverse spin polarization of the gas.

\ack
The authors acknowledge support by the University of the Philippines OVPAA through Grant number OVPAA-BPhD-2012-05. 

\section*{References}
\providecommand{\newblock}{}

\end{document}